\documentclass[%
reprint,
superscriptaddress,
amsmath,amssymb,
aps,
pra,
floatfix,
longbibliography
]{revtex4-1}

\usepackage{graphicx}
\usepackage{dcolumn}
\usepackage{bm}
\usepackage{braket}
\usepackage{hyperref}
\usepackage{booktabs}

\usepackage[usenames,dvipsnames]{xcolor}

\newcommand{\imag}{\mathrm{i}}
\newcommand{\e}[1]{\mathrm{e}^{#1}}
\newcommand{\dd}{\mathrm{d}}

\renewcommand{\Re}{\mathrm{Re}}
\renewcommand{\Im}{\mathrm{Im}}
\newcommand{\rom}[1]{\uppercase\expandafter{\romannumeral #1\relax}}

\begin{document}

\preprint{APS/123-QED}

\title{Transient spectroscopy from time-dependent electronic-structure theory without multipole expansions}

\author{Einar Aurbakken}
\email{einar.aurbakken@kjemi.uio.no}
\affiliation{Hylleraas Centre for Quantum Molecular Sciences,
Department of Chemistry, University of Oslo, Norway
}%

\author{Benedicte Sverdrup Ofstad}
\affiliation{Hylleraas Centre for Quantum Molecular Sciences,
Department of Chemistry, University of Oslo, Norway
}%

\author{H{\aa}kon Emil Kristiansen}
\affiliation{Hylleraas Centre for Quantum Molecular Sciences,
Department of Chemistry, University of Oslo, Norway
}%

\author{{\O}yvind Sigmundson Sch{\o}yen}
\affiliation{Department of Physics,
University of Oslo, Norway
}%

\author{Simen Kvaal}
\affiliation{Hylleraas Centre for Quantum Molecular Sciences,
Department of Chemistry, University of Oslo, Norway
}%

\author{Lasse Kragh S{\o}rensen}
\affiliation{University Library, University of Southern Denmark, DK-5230 Odense M, Denmark}%

\author{Roland Lindh}
\affiliation{Department of Chemistry---BMC,
Uppsala University, Sweden
}%

\author{Thomas Bondo Pedersen}
\email{t.b.pedersen@kjemi.uio.no}
\affiliation{Hylleraas Centre for Quantum Molecular Sciences,
Department of Chemistry, University of Oslo, Norway
}

\date{\today}

\begin{abstract}
Based on the work done by an electromagnetic field on an atomic or molecular electronic system,
a general gauge invariant formulation of transient absorption spectroscopy is presented within the semi-classical approximation.
Avoiding multipole expansions, a computationally viable expression for the spectral response function 
is derived from the minimal-coupling Hamiltonian of an electronic system interacting with one or more
laser pulses described by a source-free, enveloped electromagnetic vector potential.
With a fixed-basis expansion of the electronic wave function, the computational cost of
simulations of laser-driven electron dynamics beyond the dipole approximation is the same
as simulations adopting the dipole approximation.
We illustrate the theory by time-dependent configuration interaction and coupled-cluster simulations of
core-level absorption and circular dichroism spectra.
\end{abstract}

\maketitle


\section{\label{sec:level1}Introduction}

Using technology developed in the past two decades, ultrashort laser pulses with attosecond duration have enabled the observation and manipulation of
multi-electron dynamics in atoms, molecules, and materials, thus opening new research avenues in physics and chemistry \cite{Corkum2007,Krausz2009,Nisoli2017}.
Quantum-mechanical simulations are mandatory to properly understand, interpret, and predict advanced attosecond experiments.
While nuclear motion becomes important on longer time-scales (femtoseconds),
one- and multi-electron ionization dynamics constitute major challenges for time-dependent electronic-structure simulations, along with electron-correlation effects \cite{klinker_electron_2018}.

Single active electron (SAE) models \cite{kulander_time-dependent_1988,schafer_energy_1990,kulander_dynamics_1993}
that, at best, only account for electron correlation through an effective potential are widely used to study processes
induced by lasers with frequency well below any multi-electron excitation energy. As the frequency increases and approaches 
resonance with a multi-electron excited state, the SAE approximation breaks down and a correlated many-body method should
be applied instead \cite{lezius_nonadiabatic_2001,markevitch_nonadiabatic_2003}.

Regardless whether the SAE model or a many-body description is used, most simulations of laser-induced processes employ the
electric-dipole approximation where the magnetic component of the laser field is neglected and the electric component is
assumed to be spatially uniform. This is an excellent approximation when the spatial extent of the electronic system is small compared
with the wavelength of the laser field.
Attosecond laser pulses, however, are commonly generated by high harmonic generation in the extreme ultraviolet and X-ray spectral regions
where beyond-dipole effects may become non-negligible.
It is, therefore, of interest to include higher-order electric and magnetic multipole interactions in simulations of laser-driven
electron dynamics, preferably without incurring a significant computational penalty.

Within response theory \cite{Olsen1985}, which is essentially time-dependent perturbation theory Fourier-transformed to the frequency domain,
beyond-dipole effects have been studied using the full plane-wave vector potential for the semiclassical
description of the matter-field interaction \cite{ListNannaHolmgaard2015Btea,ListNannaHolmgaard2017Rala,SorensenLasseKragh2019Ioas,vanHornMartin2022Pcat}.
Due to the use of perturbation theory and the neglect of terms quadratic in the vector potential,
these studies are limited to weak laser fields but do not suffer from issues such as origin-dependence and slow basis-set
convergence that may arise from the use of multipole expansions \cite{Bernadotte2012,Lestrange2015,sorensen_gauge_2017,van_horn_transition_2023}.
Conceptually, at least, it is rather straightforward to generalize the response-theory approaches
to the time domain, avoiding perturbation theory altogether and hence enabling the study of both weak- and strong-field processes
without multipole expansions.

The theory of transient absorption spectroscopy (TAS), see, e.g., recent work by \citet{WuMengxi2016Tosa}, has been formulated in the framework of
the electric-dipole approximation. In the present work, we present a generalization that accounts for the presence of spatially non-uniform fields, which
reduces to the original formulation in the long-wavelength (electric-dipole) limit. In line with the previous work based on response
theory \cite{ListNannaHolmgaard2015Btea,ListNannaHolmgaard2017Rala,SorensenLasseKragh2019Ioas,vanHornMartin2022Pcat},
we present initial test simulations on small molecules in the weak-field limit using time-dependent configuration-interaction
(TDCI) \cite{Rohringer2006,krause_molecular_2007,schlegel_electronic_2007,pabst_decoherence_2011,white_computation_2016}
and time-dependent coupled-cluster (TDCC) \cite{Ofstad2023} theories.
Ignoring ionization processes, we use static, atom-centered Gaussian basis sets such that the prerequisite integrals involving the
full plane-wave vector potential can be computed using the recent implementation reported by \citet{SorensenLasseKragh2019Ioas}.
This allows us to validate our implementation of the generalized theory of TAS by comparing with previously reported theoretical pump-probe
and X-ray absorption spectra. In addition, we compute the anisotropic X-ray circular dichroism (CD) spectrum of hydrogen peroxide generated from
simulations of the electrons interacting with circularly polarized laser
pulses \cite{KfirOfer2015Gobp,BandraukAndreD.2016Cpap,ShaoRenzhi2020Goic,MaGuangjin2016Icpa,ZhongConglin2021Poii,ChenCong2016Troc},
comparing with the CD spectrum predicted by the rotatory strength tensor \cite{PEDERSENTB1995ACAD}.

\section{Theory}

Atomic and molecular transient (as well as steady-state) absorption spectra can be obtained by computing the spectral response function
$S(\omega)$ which, in turn, is obtained from a frequency-resolved analysis of the total energy transfer $\Delta \mathcal{E}$ between an electromagnetic field
and the electronic  system. The spectral response function $S(\omega)$ is defined such that it satisfies the relation
\begin{equation}\label{eq:S}
    \Delta \mathcal{E} = \int_0^\infty \dd\omega\ \omega S(\omega).
\end{equation}
The absorption cross section $\sigma(\omega)$ can be computed as
\begin{equation}
    \sigma(\omega) = \frac{\omega S(\omega)}{I(\omega)},
\end{equation}
where $I(\omega)$ is the total field energy per unit area at frequency $\omega$.
In this work, however, we shall focus on the spectral response function.
We first formulate a general, gauge invariant theory for the energy transfer, proceeding to the derivation of the spectral response function
for the specific case of an enveloped, source-free electromagnetic field without multipole expansion.

\subsection{Energy transfer}

We consider an atomic or molecular electronic system exposed to
the classical electromagnetic fields
\begin{align}
    \label{eq:Efield}
    &\bm{E}(\bm{r},t) = -\partial_t\bm{A}(\bm{r},t) - \nabla \phi(\bm{r},t), \\
    \label{eq:Bfield}
    &\bm{B}(\bm{r},t) = \nabla \times \bm{A}(\bm{r},t),
\end{align}
where $\bm{A}(\bm{r},t)$ and $\phi(\bm{r},t)$ are the vector and scalar potentials, respectively.
Specifically, we will consider the interactions of the electrons with laser pulses, i.e.,
the physical electric and magnetic fields, $\bm{E}$ and $\bm{B}$, are nonzero only in a finite time interval
and vanish as $t\to\pm\infty$.
Within the nonrelativistic, clamped-nuclei Born-Oppenheimer approximation the time evolution of the electronic 
system is governed by the electronic Schr{\"o}dinger equation
\begin{equation}\label{eq:TDSE}
    \imag\ket{\dot{\Psi}(t)} = H(t) \ket{\Psi(t)}, \qquad \ket{\Psi(t\to -\infty)} = \ket{\Psi_0},
\end{equation}
where $\ket{\Psi_0}$ is the initial wave function of the electrons, typically the ground-state wave function in the absence
of external fields.
The semiclassical, minimal-coupling Hamiltonian is given by
\begin{equation}
    H(t) = \frac{1}{2}\pi^2(\bm{r},t) + W - \phi(\bm{r},t),
\end{equation}
where $\bm{\pi}(\bm{r},t) = \bm{p} + \bm{A}(\bm{r},t)$ is the kinetic momentum operator
and $W$ represents all Coulomb interactions among the electrons and (clamped) nuclei.
Throughout this paper, summation over electrons will be implicitly assumed for brevity of notation, and Hartree atomic units are used.
We have also skipped the spin-Zeeman term as we will use only closed-shell, spin-restricted wave functions in the present
work.

We wish to derive a general expression for the spectral response function $S(\omega)$ in Eq.~\eqref{eq:S}.
Physically, the total energy transfer $\Delta \mathcal{E}$
expresses the work performed on the electronic system by the external electromagnetic fields,
and the rate of change of the energy is referred to as the power.
In classical electrodynamics~\cite{Jackson}, the power function of an electron in an electromagnetic field is given by
$P = -\bm{E}\cdot\bm{v}$, where $\bm{v}$ is the velocity of the electron.
This is also the energy lost by the electromagnetic field as calculated by Poynting's theorem~\cite{Jackson},
ensuring energy conservation (of the particle and field systems together).
The quantum-mechanical power operator can be obtained by Weyl quantization~\cite{Weyl1927}
as
\begin{equation}\label{eq:power_operator}
    P(\bm{r}, t) = -\frac{1}{2}\left(\bm{E}(\bm{r},t)\cdot\bm{\pi}(\bm{r},t) + \bm{\pi}(\bm{r},t)\cdot\bm{E}(\bm{r},t)\right).
\end{equation}
Hence, we may express the total energy transferred from the field to the electronic system as
\begin{equation}\label{eq:general_dE}
    \Delta \mathcal{E} = \int_{-\infty}^{\infty}\dd t\ \braket{P(\bm{r},t)}.
\end{equation}

In previous work on transient absorption spectroscopy---see,
e.g., Refs.~\cite{TannorDavid2007Itqm,WuMengxi2016Tosa,SkeidsvollAndreasS2020Tctf,Guandalini2021}---the energy
transfer is expressed as the integral
\begin{equation}\label{eq:dE_ddt<E>}
    \Delta \mathcal{E} = \int_{-\infty}^{\infty} \dd t\ \frac{\dd \mathcal{E}(t)}{\dd t},
\end{equation}
where $\mathcal{E}(t)$ is the instantaneous energy of the electrons.
At this point, the instantaneous energy is typically equated with the quantum-mechanical expectation value of the Hamiltonian,
$\braket{H(t)} = \braket{\Psi(t) \vert H(t) \vert \Psi(t)}$. In general, however,
neither the expectation value $\braket{H(t)}$ nor the Hamilton function in classical mechanics \cite{Goldstein}
equals the energy of the electrons when a time-dependent external electromagnetic field is present.
This is clear from the fact that both $\braket{H(t)}$ and $\dd\braket{H(t)}/\dd t$ are gauge-dependent quantities.
Instead, the operator~\cite{YangKuo-Ho1976Gtaq,KobeDH1987Eoac}
\begin{equation}\label{eq:energy_operator}
    K(t) = H(t) + \phi(\bm{r},t) = \frac{1}{2}\pi^2(\bm{r},t) + W,
\end{equation}
can be regarded as a (generally time-dependent) energy operator which yields gauge invariant expansion coefficients
and transition probabilities when the wave function is expanded in its (generally time-dependent) eigenstates.
Using the energy operator, Eq.~\eqref{eq:energy_operator}, and the Ehrenfest theorem, we find
\begin{equation}
    \frac{\dd \mathcal{E}(t)}{\dd t} = \frac{\dd \braket{K(t)}}{\dd t}
    = \braket{P(\bm{r},t)},
\end{equation}
which leads to Eq.~\eqref{eq:general_dE} upon substitution in Eq.~\eqref{eq:dE_ddt<E>}.
We refer to references \cite{YangKuo-Ho1976Gtaq,YangKuo-Ho1976Gtaq2,KobeDonaldH.1978Gifo,KobeDonaldH.1978Iogi,KobeDH1980Gnlo,Aharonov1981,KobeDH1982Giiq,KobeDH1987Eoac} for further discussions of the intricacies of gauge invariance in external time-varying fields.

Within the electric-dipole approximation, $\bm{A}(\bm{r},t) \approx \bm{A}(\bm{0},t) = \bm{A}(t), \quad \phi(\bm{r},t) = 0$, which was assumed in 
previous work \cite{TannorDavid2007Itqm,WuMengxi2016Tosa,SkeidsvollAndreasS2020Tctf,Guandalini2021}, the power operator becomes
$P(t) = -\bm{\pi}(t)\cdot\bm{E}(t)$. Inserting this expression into Eq.~\eqref{eq:general_dE} yields
\begin{equation}
    \Delta \mathcal{E} = -\int_{-\infty}^{\infty}\dd t\ \braket{\bm{\pi}(t)} \cdot \bm{E}(t).
\end{equation}
Using the Ehrenfest theorem,
\begin{equation}\label{eq:Ehrenfest_r_pi}
    \frac{\dd \braket{\bm{r}}}{\dd t} = \braket{\bm{\pi}(t)},
\end{equation}
and integration by parts, we arrive at
\begin{equation}
    \Delta \mathcal{E} = \int_{-\infty}^{\infty}\dd t\ \braket{\bm{r}} \cdot \dot{\bm{E}}(t),
\end{equation}
which agrees with the expressions obtained in Refs.~\cite{TannorDavid2007Itqm,WuMengxi2016Tosa,SkeidsvollAndreasS2020Tctf,Guandalini2021}.

Identifying the instantaneous energy as the expectation value $\braket{H(t)}$ is valid when the scalar potential vanishes
which, in turn, is a valid choice with the Coulomb gauge condition $\nabla \cdot \bm{A}(\bm{r},t) = 0$ whenever the electric field is divergence-free (no charge contributions to the electric field), i.e. within radiation gauge \cite{schlicher1984interaction}.
It is a peculiarity of the electric-dipole approximation that the correct energy transfer is obtained from
$\braket{H(t)}$ with the choices $\bm{A}(\bm{r},t) = \bm{0}$ and $\phi(\bm{r},t) = -\bm{r}\cdot\bm{E}(t)$.

\subsection{Representation of laser pulses without multipole expansion}

From here on we will assume a divergence-free electric field and work in the radiation gauge such that $K(t) = H(t)$.
Following common practice, we separate the Hamiltonian into a time-independent and a time-dependent part,
\begin{align}
    &H(t) = H_0 + V(t), \\
    &H_0 = \frac{1}{2}p^2 + W, \\
    &V(t) = \bm{A}(\bm{r},t)\cdot \bm{p} + \frac{1}{2}A^2(\bm{r},t).
\end{align}
In the context of time-dependent perturbation theory or frequency-dependent response theory,
the weak-field approximation---i.e., neglecting the term quadratic in the vector potential---is
usually invoked, although it is not formally necessary to do
so \cite{ListNannaHolmgaard2015Btea,ListNannaHolmgaard2017Rala,SorensenLasseKragh2019Ioas,vanHornMartin2022Pcat}.
For the real-time simulations pursued in the present work, invoking the weak-field approximation does not lead
to any simplifications and, hence, we retain the quadratic term in all simulations.

The vector potential that solves the Maxwell equations within the Coulomb gauge is a linear combination of plane waves. However, this is impractical for modelling ultra-fast laser pulses. 
We will instead model the vector potential as a linear combination of enveloped plane waves
\begin{align}\label{eq:vec_pot_expansion}
    &\bm{A}(\bm{r},t) = \sum_{m}\bm{A}_m(\bm{r},t) G_m(t)\nonumber \\
    &= \sum_{m}A_m \Re \{ \bm{u}_m \e{\imag (\bm{k}_m \cdot \bm{r} - \omega_m t - \gamma_m)} \}G_m(t),
\end{align}
where each term in the sum models a single pulse with amplitude $A_m$,
carrier frequency $\omega_m$,
and carrier-envelope phase $\gamma_m$.
The Coulomb gauge condition implies that
the (complex) polarization vector $\bm{u}_m$ is orthogonal to the real wave vector $\bm{k}_m$,
which has length $\omega_m/c$ where $c$ is the speed of light.
The electric- and magnetic-field amplitudes of each pulse are $E_m = \omega_m A_m$ and
$B_m = E_m/c$, respectively, and we define the peak intensity of each pulse to be
\begin{equation}
    I_m = \frac{1}{2}\epsilon_0 c E_m^2.
\end{equation}
Chirped laser pulses can be modelled by letting $\gamma_m$ be time-dependent.

In experimental work, Gaussian functions are often favored for the envelopes $G_m(t)$. In numerical studies, however, 
Gaussians are inconvenient due to their long tails and infinite support.
For this reason, we use trigonometric envelopes on the form~\cite{BarthI2009Tpef}
\begin{equation}\label{eq:envelope}
     G_m(t) =
     \begin{cases}
         \cos^n{\left(\frac{\pi (t-t_m)}{T_{mn}}\right)} & \vert t-t_m \vert \leq \frac{T_{mn}}{2} \\
         0 & \vert t-t_m \vert > \frac{T_{mn}}{2}
     \end{cases}
\end{equation}
where $n > 0$ is a chosen parameter, $t_m$ is the central time of pulse $m$, and $T_{mn}$ is the total duration
of $\bm{A}_m$.
The total duration depends on $n$ and may be computed from
\begin{equation}
    T_{mn} = \frac{\pi \tau_m}{2\arccos(2^{-1/(2n)})},
\end{equation}
where $\tau_m$ is the full width at half maximum of $G^2_m(t)$, i.e., $\tau_m$ is approximately the desired
experimental pulse duration defined from the intensity distribution \cite{BarthI2009Tpef}.

The trigonometric envelopes, Eq.~\eqref{eq:envelope}, define a sequence of functions that rapidly
and uniformly converges to the Gaussian function $\exp(-2 \ln{(2)} (t-t_m)^2/\tau_m^2)$ for increasing values of $n$ \cite{BarthI2009Tpef}.
Moreover, in contrast to finite numerical representations of Gaussian envelopes,
the trigonometric envelopes guarantee that the dc (zero-frequency) component of the electric field vanishes identically
for any choice of $n > 0$, in agreement with the far-field approximation of the Maxwell equations \cite{Rauch2006}.

A similar setup has been used before in grid treatments of single-electron systems~\cite{PhysRevLett.95.043601,PhysRevLett.97.043601,MoeThoreEspedal2018Ioah} where pulses on the form
\begin{align}\label{eq:vector_potential_spatial_envelope}
    \bm{A}(\bm{r},t) = A_0 \sin^2{\left(\frac{\pi (\omega t - \bm{k}\cdot\bm{r})}{\omega T} \right)}\sin{\left(\omega t - \bm{k}\cdot\bm{r}\right)} \bm{u},
\end{align}
were used. Here, $\bm{u}$ is a real polarization vector and the envelope depends \emph{both} on time and on spatial coordinates.
This has the benefit of modelling the overall shape of the pulse in space, albeit with potential edge effects if the approximation
$\bm{A}(\bm{r},t) \approx 0$ at $t=0$ and $t=T$ is made along with a neglect of the spatial non-periodicity.
The pulse with the purely time-dependent envelope, Eq.~\eqref{eq:envelope} with $n=2$, may be regained from the spatio-temporal envelope 
by an expansion through lowest order in $\bm{k}\cdot\bm{r}/n_{cyc}$, where $n_{cyc}$ is the number of optical cycles of the pulse.

\subsection{The spectral response function}

Since we have assumed a divergence-free electric field, the power operator becomes
$P(\bm{r},t) = -\bm{E}(\bm{r},t)\cdot\bm{\pi}(\bm{r},t)$, and Eq.~\eqref{eq:general_dE} simplifies to
\begin{align}\label{eq:energy_transfer}
    \Delta \mathcal{E} = -\int_{-\infty}^{\infty}\dd t\ \langle \bm{E}(\bm{r},t)\cdot \bm{\pi}(\bm{r},t)\rangle.
\end{align}
Using the Fourier transform convention
\begin{align}
    &f(t) = \mathcal{F}_\omega\bigl[\tilde{f}(\omega)\bigr] = \frac{1}{\sqrt{2\pi}}\int_{-\infty}^{\infty}\dd \omega\ \tilde{f}(\omega)e^{\imag\omega t}, \\
    &\tilde{f}(\omega) = \mathcal{F}_t\bigl[f(t)\bigr] = \frac{1}{\sqrt{2\pi}}\int_{-\infty}^{\infty}\dd t\ f(t)e^{-\imag\omega t},
\end{align}
the integration over time in Eq.~\eqref{eq:energy_transfer} can be turned into an integration over frequency,
\begin{equation}
    \Delta \mathcal{E} = \int_{-\infty}^\infty \dd\omega\ Y(\omega),
\end{equation}
with
\begin{equation}\label{eq:Yw}
    Y(\omega) = - \mathcal{F}_t \left[ \imag\omega \langle \Tilde{\bm{A}}(\bm{r},\omega)^*\cdot\bm{\pi}(\bm{r},t) \rangle\right],
\end{equation}
where we have used $\Tilde{\bm{E}}(\bm{r},\omega)^* = \imag\omega\Tilde{\bm{A}}(\bm{r},\omega)^*$. 

Introducing
\begin{align}\label{eq:definitions}
    &f_{1,m}(\bm{r}) = \cos{(\bm{k}_m\cdot\bm{r})},\\&f_{2,m}(\bm{r}) = \sin{(\bm{k}_m\cdot\bm{r})},\\& g_{1,m}(t) = \cos{(\omega_mt+\gamma_m)}G_m(t),
    \\ &g_{2,m}(t) = \sin{(\omega_mt+\gamma_m)}G_m(t),\\
    &\bm{u}_m^{ij} = \delta_{ij} \Re(\bm{u}_m) + \epsilon_{ij} \Im(\bm{u}_m),
\end{align}
where $\delta_{ij}$ is the Kronecker delta and $\epsilon_{ij}$ is the Levi-Civita symbol, the vector potential, Eq.~\eqref{eq:vec_pot_expansion},
can be recast as
\begin{equation}\label{eq:Afull_expansion}
    \bm{A}_m(\bm{r},t)G_m(t) = A_m\sum_{i,j=1}^2\bm{u}_m^{ij}f_{i,m}(\bm{r})g_{j,m}(t).
\end{equation}
Equation \eqref{eq:Yw} can now be written as
\begin{equation}
    Y(\omega) =  \omega \sum_{m}\sum_{i,j=1}^2  \Tilde{F}_{ij,m}(\omega) \Tilde{g}_{j,m}(-\omega),
\end{equation}
where $\Tilde{F}_{ij,m}(\omega)$ is the Fourier transform of the function
\begin{align}\label{eq:F}
    &F_{ij,m}(t) = -\imag A_m\bm{u}_m^{ij}\cdot \Biggl[ \langle f_{i,m}(\bm{r})\bm{p}\rangle \Biggr. \nonumber\\&\left.+ \sum_{n}\sum_{k,l=1}^2 A_n \bm{u}_n^{kl}\langle f_{k,n}(\bm{r}) f_{i,m}(\bm{r}) \rangle g_{l,n}(t)\right].
\end{align}
Hence,
\begin{equation}
    \Delta \mathcal{E} = \int_{0}^{\infty}\dd\omega\ \omega \sum_{m}\sum_{i,j=1}^2 \left[(1 - \mathcal{P}) \Tilde{F}_{ij,m}(\omega) \Tilde{g}_{j,m}(-\omega) \right],
\end{equation}
where $\mathcal{P}$ is the parity operator defined by $\mathcal{P}f(\omega) = f(-\omega)$.
The spectral response function thus becomes
\begin{equation}\label{eq:general_S}
    S(\omega) =  \sum_{m}\sum_{ij=1}^2 (1 - \mathcal{P})\Tilde{F}_{ij,m}(\omega) \Tilde{g}_{j,m}(-\omega),
\end{equation}
which can be computed by sampling $F_{ij,m}(t)$ during a simulation, followed by Fourier transformation in a post-processing step.

In the electric-dipole approximation, $f_{1,m}(\bm{r}) = 1$ and $f_{2,m}(\bm{r}) = 0$, and in this case the spectral response function reduces to 
\begin{equation}\label{eq:pia_S}
    S(\omega) = 2 \Im \left[ \langle\Tilde{\bm{\pi}}\rangle(\omega) \cdot \Tilde{\bm{A}}(\omega)^* \right],
\end{equation}
or, equivalently,
\begin{align}\label{eq:de_S}
    S(\omega) = -2 \Im \left[ \langle\Tilde{\bm{d}}\rangle(\omega) \cdot \Tilde{\bm{E}}(\omega)^* \right],
\end{align}
in terms of the dipole operator $\bm{d}=-\bm{r}$. The latter expression, Eq.~\eqref{eq:de_S}, was used in
Refs.~\cite{TannorDavid2007Itqm,WuMengxi2016Tosa,SkeidsvollAndreasS2020Tctf,Guandalini2021}.
We remark that Eqs.~\eqref{eq:pia_S} and \eqref{eq:de_S} are equivalent only if the Ehrenfest theorem, Eq.~\eqref{eq:Ehrenfest_r_pi},
is satisfied, i.e., for fully variational many-body wave function approximations, and in the limit of complete one-electron basis set.

For the visual presentation of spectra we use normalized spectral response functions
\begin{align}\label{eq:normalized_S}
    \Bar{S}(\omega) = \frac{S(\omega)}{\max(S_\text{ref}(\omega))}
\end{align}
where $S_\text{ref}$ is the spectral response function of some reference system.

\section{Numerical experiments}

In order to test the multipole-expansion-free theory outlined above, we will investigate the following aspects:
\begin{enumerate}
    \item Reproducibility of results obtained within the electric-dipole approximation in the long wavelength limit:
        Core-level pump-probe spectrum of LiH (section \ref{subsec:LiH_pump-probe}).
    \item Reproducibility of results obtained with low-order multipole expansions for short wavelengths:
        Pre-K-edge quadrupole transitions in Ti (section \ref{subsec:Pre_K-edge_quadrupole_transitions}).
    \item Intrinsically beyond-dipole phenomena:
        Anisotropic circular dichroism (section \ref{subsec:Circular_dichroism}).
\end{enumerate}

\subsection{Computational details}\label{subsec:Computational_details}

All simulations are performed with the open-source software Hylleraas Quantum Dynamics (HyQD)~\cite{HyQD}.
We employ a series of nonrelativistic, closed-shell, spin-restricted time-dependent electronic-structure
methods based on a single reference Slater determinant
built from spin orbitals expanded in a fixed atom-centered Gaussian basis set. The orbital expansion coefficients are either kept constant (static orbitals)
at the ground-state Hartree-Fock (HF) level or allowed to vary in response to the external field (dynamic orbitals). Static orbitals are
used in the time-dependent configuration interaction singles (TDCIS)~\cite{LiXiaosong2020RTES},
time-dependent second-order approximate coupled-cluster singles-and-doubles (TDCC2)~\cite{ChristiansenOve1995Tsac,KristiansenHakonEmil2022LaNO},
and time-dependent coupled-cluster singles-and-doubles (TDCCSD)~\cite{PedersenThomasBondo2019Siap} methods. Dynamic orbitals are
used in the time-dependent Hartree-Fock (TDHF)~\cite{LiXiaosong2020RTES},
time-dependent orbital-optimized second-order M{\o}ller-Plesset (TDOMP2)~\cite{KristiansenHakonEmil2022LaNO,PathakHimadri2020Tocm},
and orbital-adaptive time-dependent coupled-cluster doubles (OATDCCD)~\cite{KvaalSimen2012Aiqd} methods.
Only the methods using dynamic orbitals are gauge invariant (in the limit of complete basis set)~\cite{Pedersen1997,Pedersen1998,Pedersen1999,Pedersen2001}.
No splitting of the orbital space is used in the OATDCCD method which, therefore, is identical to the nonorthogonal orbital-optimized
coupled-cluster doubles model \cite{Pedersen2001}.
In the TDHF and TDOMP2 models the dynamic-orbital evolution is constrained to maintain orthonormality throughout, whereas in OATDCCD theory the
dynamic orbitals are biorthonormal~\cite{KvaalSimen2012Aiqd,Pedersen2001}.

The methods can be roughly divided into three approximation levels.
The TDCIS and TDHF methods are the least computationally demanding ones (with formal scaling $\mathcal{O}(K^4)$ with $K$ the number of basis functions)
and do not account for electron correlation.
The TDCCSD and OATDCCD methods are the most accurate and most expensive ($\mathcal{O}(K^6)$) methods with full treatment of double excitations.
Finally, the TDCC2 and TDOMP2 methods are intermediate in both accuracy and computational cost ($\mathcal{O}(K^5)$).
The TDCC2 method is a second-order approximation to the TDCCSD model, while the TDOMP2 model is the analogous second-order approximation to the 
orbital-optimized coupled-cluster doubles model~\cite{Pedersen1999,SatoTakeshi2018CToc}.
The doubles treatment of TDOMP2 theory is essentially identical to that of TDCC2 theory but provides full orbital relaxation through
unitary orbital rotations instead of the singles excitations of static-orbital coupled-cluster theory.

Since fixed, atom-centered Gaussian basis sets are used, ionization cannot be described and, therefore, the simulations are
restricted to weak electromagnetic field strengths.
On the other hand, the fixed basis set allows us to compute matrix elements of the plane-wave interaction operators using
the OpenMolcas software package~\cite{FdezGalvanIgnacio2019OFSC,Aquilante2020} via a Python interface implemented in 
the Dalton Project~\cite{Olsen2020}.
The remainder of the Hamiltonian matrix elements 
and the ground-state HF orbitals are computed using the PySCF program~\cite{SunQiming2018PtPs} with the exception of the LiH system for which
the Dalton quantum chemistry package \cite{Aidas2014} was used.
The convergence tolerance for the HF ground states is set to $10^{-10}\,\text{a.u.}$ for both the HF energy and the norm of the orbital gradients in the PySCF calculations, while the default value of $10^{-6}\,\text{a.u.}$ on the HF energy was used in the Dalton calculations.
The basis sets were obtained from the Python library \emph{Basis Set Exchange}~\cite{pritchard2019a}.
The systems are initially in the ground state which is calculated with ground-state solvers implemented in HyQD
for all the methods except the TDHF and TDCIS models, for which the ground-state wave function is computed using PySCF.
A convergence tolerance of $10^{-10}$ is also used for the amplitude residuals in the ground-state coupled-cluster calculations.

The integration of the equations of motion is done with the symplectic Gauss-Legendre integrator~\cite{PedersenThomasBondo2019Siap,HairerLubichWanner_GNI}
of order six and with a convergence threshold on the residual norm of 10$^{-10}$ for the implicit equations.
The simulations are performed with the pulse defined in Eq.~\eqref{eq:vec_pot_expansion}. The laser pulse parameters
will be given for each system below.

In actual simulations, time-dependent functions such as $F_{ij,m}(t)$ and $g_{i,m}(t)$ are computed as
discrete time series, forcing us to use the fast Fourier transform algorithm. 
To reduce the appearance of broad oscillations around the peaks due to spectral leakage,
we roughly follow the procedure used by \citeauthor{SkeidsvollAndreasS2020Tctf}~\cite{SkeidsvollAndreasS2020Tctf}.
The simulation is started at time $t<0$ when the first pulse is switched on and continued until
time $t_\text{max} > 0$ after the last pulse is switched off.
We then extend the recorded time series such that $t_\text{min}=-t_\text{max}$ to obtain a symmetric time range
about $t=0$. To do so, we use that $\bm{A}(\bm{r},t)=\bm{0}$ and hence $V(t) = 0$ in the time interval before the pulse is switched on.
We then multiply the resulting time series defined on the uniformly discretized time interval from $t_\text{min}$ to $t_\text{max}$
with the Hann function,
\begin{align}
    w_H(t) = \cos^2{\left(\frac{\pi t}{2t_\text{max}}\right)},
\end{align}
before the fast Fourier transform is performed.

\subsection{Core-level pump-probe spectrum of LiH}\label{subsec:LiH_pump-probe}

The most common experimental methods for spectral analysis of attosecond interactions employ pump-probe setups.
Therefore, we start by simulating a pump-probe spectrum for LiH.
The K pre-edge features of Li are expected at less than $60\,\text{eV}$, corresponding to a wavelength of 
$\sim\,200\,\text{{\AA}}$. In the weak-field limit, the beyond-dipole effects are expected to be quite small, allowing us to compare
with the TDCCSD simulations carried out within the electric-dipole approximation
by \citeauthor{SkeidsvollAndreasS2020Tctf}~\cite{SkeidsvollAndreasS2020Tctf}.

For the most part we follow the setup of \citeauthor{SkeidsvollAndreasS2020Tctf}~\cite{SkeidsvollAndreasS2020Tctf}.
We start the TDCCSD simulations at $t=-200\,\text{a.u.}$ and end it at $t_\text{max}=5000\,\text{a.u.}$ The pump pulse centered at $t_1=-40\,\text{a.u.}$ has
a carrier frequency of $3.55247\,\text{eV}$ and maximum electric field strength of $0.01\,\text{a.u.}$
(corresponding to a peak intensity of $3.51\times 10^{12}\,\text{W/cm}^2$), while the probe pulse centered at $t_2=0\,\text{a.u.}$ has a carrier frequency of
$57.6527\,\text{eV}$ and maximum electric field strength of $0.1\,\text{a.u.}$ (peak intensity $3.51\times 10^{14}\,\text{W/cm}^2$). Both pulses are linearly polarized in the $z$-direction (parallel to the molecular axis)
with zero carrier-envelope phases, and the propagation direction is along the $x$-axis. The beyond-dipole spectrum was generated using Eq. (\ref{eq:general_S}) while Eq. (\ref{eq:pia_S}) was used to generate the dipole spectrum. The dipole simulation was done in velocity gauge to eliminate any gauge differences between the two simulations.
We note in passing that the intensity of the probe pulse is too strong to warrant the complete neglect of ionization processes, but to facilitate
comparison with the spectra reported in Ref.~\citenum{SkeidsvollAndreasS2020Tctf} we choose to keep it.

\citeauthor{SkeidsvollAndreasS2020Tctf} used a Gaussian envelope on the electric field with root-mean-square width $\sigma_1=20\,\text{a.u.}$ for the pump pulse and
$\sigma_2=10\,\text{a.u.}$ for the probe pulse.
Here, we instead use the trigonometric approximation in Eq.~\eqref{eq:envelope} placed on the vector potential with
\begin{align}
    T_{mn} = \frac{\pi \sqrt{\ln{(2)}} \sigma_m}{\arccos(2^{-1/(2n)})}, \qquad m = 1,2,
\end{align}
and $n = 19$, which is the largest integer for which the pump pulse is strictly zero at $t=-200\,\text{a.u.}$ 

\begin{figure}[h]
\includegraphics[width=220 pt]{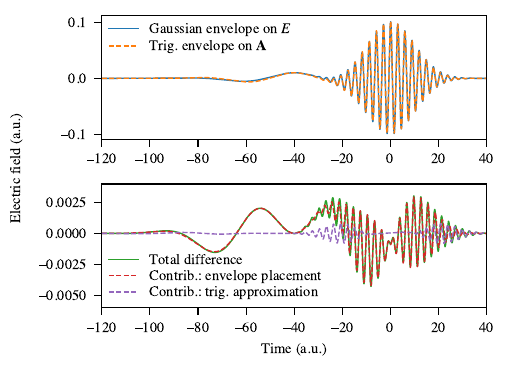}
\caption{\label{fig:pulse_comparison} Top: Electric field with a Gaussian envelope on the electric field and trigonometric envelope on the vector potential with exponent $n=19$. 
Bottom: the difference between the two pulses and contributions to it from two distinct sources.
}
\end{figure}

There are mainly three aspects of our simulations that will make our dipole spectrum different from that in Ref.~\cite{SkeidsvollAndreasS2020Tctf}:
(1) placement of a trigonometric envelope on the vector potential rather than a Gaussian envelope on the electric field,
(2) simulating in velocity gauge instead of length gauge, and
(3) using Eq. (\ref{eq:pia_S}) rather than Eq. (\ref{eq:de_S}) to generate the spectra.
The first point corresponds to effectively a different electric field component of the physical pulse.
This difference will diminish with increasing number of cycles in the pulses.
Both points (2) and (3) are due to lacking gauge invariance.
Illustrating the difference between the two pulse setups, Fig.~\ref{fig:pulse_comparison} shows the $z$-component of the electric field
at the origin, $E_z(\bm{0},t)$, with Gaussian envelope on the electric field and trigonometric envelope on the vector potential.
The bottom panel shows the difference between the two pulse setups, and the contribution to the difference due to the trigonometric
approximation and due the placement of the envelope on the vector potential rather than on the electric field.
We see that the placement of the envelope is the dominating contribution, especially in the pump region.
The difference in the pump region is also more significant due to the smaller amplitude and consequently larger relative difference.

\begin{figure}[h]
\includegraphics[width=220 pt]{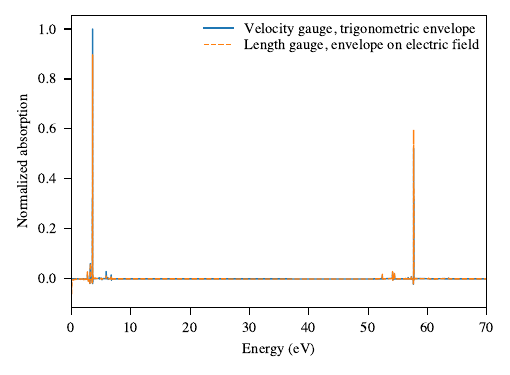}
\caption{\label{fig:len_E_gauss_vs_vel} Difference between TDCCSD pump-probe spectra computed in the electric-dipole approximation in length gauge using the Gaussian envelope placed on the electric field, and in velocity gauge with the trigonometric envelope ($n=19$) on the vector potential.}
\end{figure}

Fig.~\ref{fig:len_E_gauss_vs_vel} shows the TDCCSD dipole spectra generated with the two alternative setups.
The length-gauge spectrum is identical to that reported by \citeauthor{SkeidsvollAndreasS2020Tctf}~\cite{SkeidsvollAndreasS2020Tctf}, and although differences are
visible on the scale of the plot, we conclude that the velocity-gauge spectrum conveys the same physics.



\begin{figure*}
\includegraphics[width=440 pt]{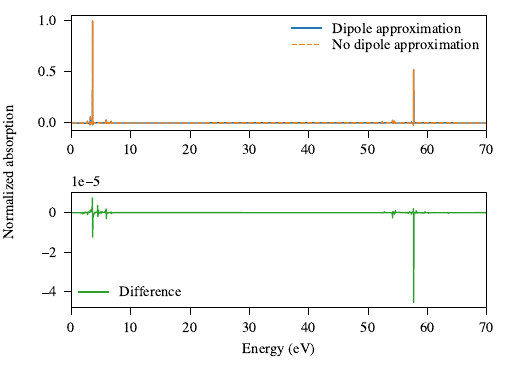}
    \caption{\label{fig:lih_vel_vs_vpi} Top: pump-probe spectrum of LiH using the TDCCSD method in the dipole approximation and with a plane-wave operator using the aug-cc-pCVDZ basis set. Bottom: the absolute difference between the spectra generated from simulations in the dipole approximation and with the plane-wave operator.
    }
\end{figure*}

Acknowledging the differences between the two setups, we will now focus on the difference between the simulations with and without the dipole approximation.
Figure \ref{fig:lih_vel_vs_vpi} compares the pump-probe spectrum simulated in the dipole approximation and with a plane-wave operator generated with equations (\ref{eq:pia_S}) and (\ref{eq:general_S}), respectively. 
Evidently, beyond-dipole effects are utterly negligible in this case: The simulation with the plane-wave operator produces a spectrum with the same 
transition frequencies as the velocity-gauge electric-dipole simulations, deviating by at most $4.5 \times 10^{-5}$, corresponding to 0.0087 \%, in relative intensity.

\subsection{K pre-edge quadrupole transitions in Ti}\label{subsec:Pre_K-edge_quadrupole_transitions}

For heavier elements, the bound core-valence excitations move up in energy and the shorter wavelengths become comparable to the ``size'' of the
atoms in terms of, e.g., covalent atomic radii. This implies that higher-order multipole effects become visible in high-resolution spectra.
The K-edge of Ti is expected at just below $5000\,\text{eV}$. This corresponds to a wavelength of roughly $2.5\,\text{\AA}$ which is comparable to
the covalent radius of Ti ($1.60\,\text{\AA}$~\cite{Cordero2008}). Consequently, one can expect visible beyond-dipole effects even in the low-intensity limit.

We consider the Ti$^{4+}$ ion and the TiCl$_{4}$ molecule. In the Ti$^{4+}$ ion the $1\text{s} \rightarrow 3\text{d}$ transition is
dipole forbidden but quadrupole allowed. In TiCl$_4$ the tetrahedral symmetry splits the $3\text{d}$-orbitals into groups of two $\text{e}$ orbitals and three $\text{t}_2$ orbitals. The $1\text{s} \rightarrow \text{e}$ transition is dipole-forbidden but quadrupole-allowed,
while the $1\text{s} \rightarrow \text{t}_2$ transition attains a dominant electric-dipole contribution due to $4\text{p}$--$3\text{d}$ mixing.
Experimentally~\cite{DeBeerGeorgeSerena2005MaLK}, a broad peak around $\!4969\,\text{eV}$ in the X-ray absorption spectrum of TiCl$_{4}$
has been assigned to the $1\text{s} \rightarrow \text{t}_2$ and $1\text{s} \rightarrow \text{e}$ with most of the intensity stemming from the former.
In the implementation presented in this paper,
electric-quadrupole and other higher-order contributions from the electromagnetic field should automatically be accounted for.

For both the Ti$^{4+}$ and TiCl$_4$ systems, we perform simulations with a $10$-cycle pulse with $n=2$ for the envelope, Eq.~\eqref{eq:envelope}, carrier frequency $181\,\text{a.u.}$ ($4925.26\,\text{eV}$),
and carrier-envelope phase $\gamma = 0$.
The duration of the simulation is $100\,\text{a.u.}$ for Ti$^{4+}$, while for TiCl$_4$ we use a total simulation time of $600\,\text{a.u.}$ to ensure
a reasonable resolution of the splitting of the d-orbitals.
The electric-field strength is $E_1 = 0.01\,\text{a.u.}$ (peak intensity $3.51\times 10^{12}\,\text{W/cm}^2$) and time step
$\Delta t = 2.5\times 10^{-4}\,\text{a.u.}$
Linearly polarized along the $x$-axis, the pulse is propagated along the $z$-axis (parallel to one of the four Ti--Cl bonds in the case of TiCl$_4$).
All Ti$^{4+}$ spectra are normalized relative to the maximum peak in the TDCCSD spectrum.

\begin{figure}[h]
\includegraphics[width=220 pt]{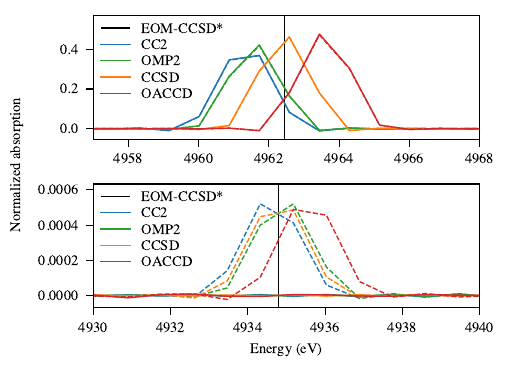}
    \caption{\label{fig:k_edge_corr_plot} Ti$^{4+}$ K pre-edge spectrum obtained from simulations with the ANO-RCC-VDZ basis set.
    Solid lines are obtained with the velocity-gauge electric-dipole approximation,
    while dashed lines are obtained with the plane-wave operator. Top: The dipole-allowed $1\text{s} \rightarrow 4\text{p}$ transition.
    Bottom: The quadrupole-allowed $1\text{s} \rightarrow 3\text{d}$ transition.
    Vertical black lines indicate the EOM-CCSD frequencies reported by \citeauthor{ParkYoungChoon2021Eomc}~\cite{ParkYoungChoon2021Eomc}. Note that, although difficult to see, the dashed lines are present also in the top panel.
    }
\end{figure}

\begin{figure}[h]
\includegraphics[width=220 pt]{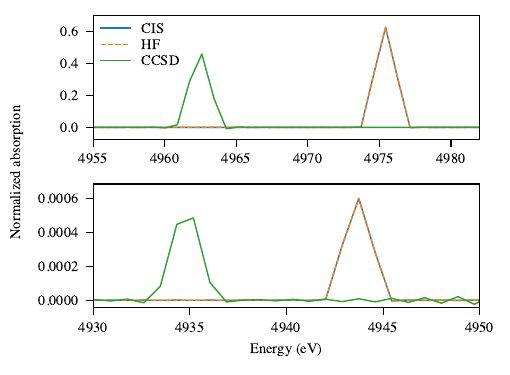}
    \caption{\label{fig:k_edge_noncorr_vs_corr}Ti$^{4+}$ K pre-edge spectrum obtained from simulations with the plane-wave operator.
    Top: The dipole-allowed $1\text{s} \rightarrow 4\text{p}$ transition.
    Bottom: The quadrupole-allowed $1\text{s} \rightarrow 3\text{d}$ transition.
    }
\end{figure}

We first consider the $1\text{s} \rightarrow 4\text{p}$ and $1\text{s} \rightarrow 3\text{d}$ transitions of Ti$^{4+}$, which have been studied recently at the
equation-of-motion coupled-cluster singles and doubles (EOM-CCSD) level of theory by
\citeauthor{ParkYoungChoon2021Eomc}~\cite{ParkYoungChoon2021Eomc} using multipole expansion up to electric octupole/magnetic quadrupole terms, for the full second-order contribution in the "mixed" length and velocity gauge \cite{sorensen_gauge_2017_2,KHAMESIAN201939,sorensen_gauge_2017}, in the framework of the Fermi golden rule.
In order to compare with their results, we use the ANO-RCC-VDZ basis set \cite{roos2005a}.
Figure \ref{fig:k_edge_corr_plot} displays the K pre-edge spectrum obtained for Ti$^{4+}$ with the TDCC2, TDOMP2, TDCCSD, and OATDCCD methods,
showing also the transition frequencies obtained by \citeauthor{ParkYoungChoon2021Eomc}~\cite{ParkYoungChoon2021Eomc}.
To within the spectral resolution of the simulation, the TDCCSD method predicts the same transition frequencies as
the static EOM-CCSD method, as expected. The intensity of the dipole-allowed $1\text{s} \rightarrow 4\text{p}$ transition
is very nearly the same both with and without the dipole approximation.
The orbital-adaptive methods yield roughly the same intensity profiles as their static-orbital counterparts, but the transition
frequencies are blue-shifted:
$\sim\!0.5\,\text{eV}$ for TDOMP2 versus TDCC2 and $\sim\!2\,\text{eV}$ for OATDCCD versus TDCCSD.
As has been observed previously \cite{KristiansenHakonEmil2022LaNO}, these blue-shifts are insignificant compared with other
sources of error such as basis-set incompleteness and higher-order correlation effects.
Electron-correlation effects are significantly more important than the orbital relaxation provided by dynamic orbitals,
as seen in Fig.~\ref{fig:k_edge_noncorr_vs_corr} where the TDCCSD spectrum is compared to the spectra obtained with the TDHF and TDCIS methods.
While the TDHF and TDCIS simulations produce virtually identical spectra, 
electron correlation causes a red-shift of the transition frequencies by roughly $8\,\text{eV}$.
The TDHF and TDCIS intensities are comparable to but slightly higher than the TDCCSD ones.
The main source of error, besides relativistic effects, is the choice of basis set: Changing from the ANO-RCC-VDZ basis set
to the cc-pVTZ basis set increases the EOM-CCSD transition frequencies by more than $28\,\text{eV}$ \cite{ParkYoungChoon2021Eomc}.

\begin{figure}[h]
\includegraphics[width=220 pt]{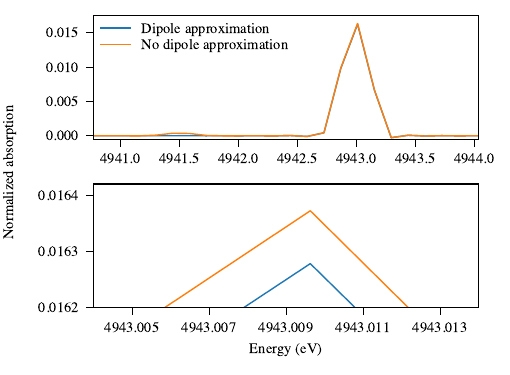}
\caption{\label{fig:ticl4} TiCl$_4$ K pre-edge spectrum from a TDCIS simulation with the plane-wave operator and with the velocity-gauge electric-dipole
    approximation. 
    Top: the dipole-forbidden $1\text{s}\rightarrow \text{e}$ transition at $4941.50\,\text{eV}$ and the
    $1\text{s} \rightarrow \text{t}_2$ transition at $4942.99\,\text{eV}$.
    Bottom: zoomed in at the $1\text{s} \rightarrow \text{t}_2$ peak.}
\end{figure}

Since we are not aiming at prediction or interpretation of experimental results in this work, we study the TiCl$_4$ K pre-edge spectrum using the
most affordable TDCIS method with the ANO-RCC-VDZ basis set. The TDCIS spectrum is shown in
Fig.~\ref{fig:ticl4}.
The dipole-forbidden $1\text{s} \rightarrow \text{e}$ transition
is visible at $4941.50\,\text{eV}$, roughly $1.5\,\text{eV}$ below the dipole-allowed $1\text{s} \rightarrow \text{t}_2$ transition at $4942.99\,\text{eV}$.
The TDCIS frequencies are blue-shifted by approximately $12\,\text{eV}$ relative to the EOM-CCSD results reported by
\citeauthor{ParkYoungChoon2021Eomc}~\cite{ParkYoungChoon2021Eomc}.
The $1\text{s} \rightarrow \text{t}_2$ transition has a
slightly higher intensity with the plane-wave interaction operator than with the dipole interaction operator. 
It should also be noted, however, that the intensities of the dipole-allowed transitions typically are 
slightly higher with the dipole approximation and, therefore, one should be careful using the dipole result as a reference for evaluating the quadrupole contribution. The deviation may be caused by a difference in the quality of the operator representation or the wave function, which may occur when propagating with different operators in a finite basis set.


\subsection{Anisotropic circular dichroism}\label{subsec:Circular_dichroism}

Circular dichroism (CD)---the difference in absorption of left and right circularly polarized radiation exhibited by chiral molecules---is
a particularly interesting case to test the implementation of the beyond-dipole interaction, since the observed effect cannot be explained within
the electric-dipole approximation. At least electric quadrupole and magnetic dipole terms must be included \cite{Rosenfeld1929,Barron2004,pecul_ab_2005,Crawford2006}
and, consequently, the differential absorption is weak compared with linear, electric-dipole absorption.
Chiroptical spectroscopies, including CD, are important for determining the absolute configuration of chiral molecules  
and core-resonant CD is particularly well suited to gauge local molecular chirality \cite{Zhang2017}.
As Eq.~\eqref{eq:general_S} was derived assuming complex polarization vectors, the implementation presented here can easily be used to
generate spectra involving pulses with circular (or, more general, elliptical) polarization, including at short wavelengths.

As alluded to above, the leading contributions to a CD spectrum arise from the magnetic-dipole and electric-quadrupole terms in the multipole expansion
of the vector potential. In an isotropic sample, the quadrupole contribution vanishes since the electric dipole--electric quadrupole component of the
rotatory strength tensor is traceless \cite{PEDERSENTB1995ACAD}. As a prototypical example which previously has been used to test new implementations of
CD spectra \cite{Rizzo2006,Friese2016,scott_electronic_2021}, we will consider the  H$_2$O$_2$ molecule in a chiral conformation
with fixed orientation relative to the external laser pulse.

The CD spectrum is calculated as the difference between the spectral response functions of two distinct simulations:
one with left circular polarization and one with right circular polarization of the pulse. We define the normalized differential absorption as
\begin{align}
    \Bar{S}_{l-r}(\omega) = \Bar{S}_l(\omega) - \Bar{S}_r(\omega)
\end{align}
where $\Bar{S}_{l}(\omega)$ and $\Bar{S}_{r}(\omega)$ are the normalized spectral response functions for the left and right circularly polarized pulses.

\begin{figure}[h]
\includegraphics[width=180 pt]{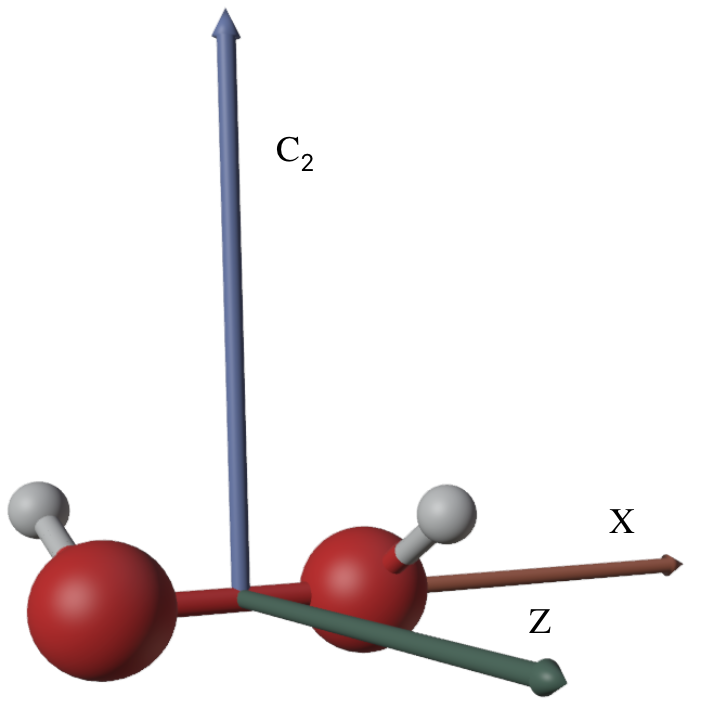}
    \caption{\label{fig:h2x2_mol}Definition of Cartesian coordinate system for H$_2$O$_2$ (C$_2$ point group).}
\end{figure}

The molecular geometry of H$_2$O$_2$, depicted in Fig.~\ref{fig:h2x2_mol} along with the Cartesian axis definitions,
is taken from Ref.~\citenum{Redington1962}. The Cartesian coordinates can be found in the supplementary material.
We choose the polarization vectors such that $\bm{u}^l + \bm{u}^r = \hat{\bm{j}}$, where $\hat{\bm{j}}$ is a unit vector aligned with the C$_2$ axis and
superscripts $r$ and $l$ refer to right and left circular polarization, respectively, as seen from the source.
We run two pairs of simulations with the propagation direction along the $x$-axis and along the $z$-axis.
For the propagation direction along the $x$-axis we use $\bm{u}^r = (0,1,\imag)$ and $\bm{u}^l = (0,1,-\imag)$,
and for the propagation direction along the $z$-axis we use $\bm{u}^{r} = (-\imag,1,0)$ and $\bm{u}^{l} = (\imag,1,0)$.
We use a carrier frequency in the K-edge region of oxygen,  $\omega = 20\,\text{a.u.}$ ($544.23\,\text{eV}$), 
and carrier-envelope phase $\gamma = 0$. The duration of the
laser pulse is $10$ optical cycles and the trigonometric envelope is defined with $n=2$, which corresponds to $\tau = 1.14\,\text{a.u.}$.
The electric-field strength is $E_1 = 0.01\,\text{a.u.}$ (peak intensity $3.51\times 10^{12}\,\text{W/cm}^2$) and the carrier-envelope phase is $\gamma = 0$.
The time step is $\Delta t = 0.005\,\text{a.u.}$ and the total simulation time is $1000\,\text{a.u.}$
We use the TDHF, TDCIS, TDCC2, TDOMP2, and TDCCSD methods with the cc-pVDZ basis set\cite{dunning1989a,woon1993a}, and
the spectra for propagation direction along the $x$- and $z$-axes are normalized with respect to the corresponding TDCIS simulation.

\begin{figure}[h]
\includegraphics[width=220 pt]{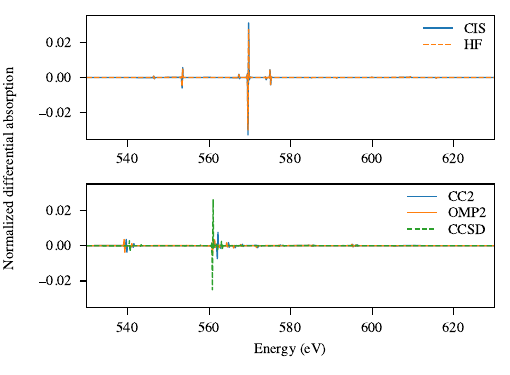}
    \caption{\label{fig:circdich_h2o2_methods_0}
    Differential spectra obtained with the cc-pVDZ basis set in the K-edge region of H$_2$O$_2$ with propagation direction along the $x$-axis.
    }
\end{figure}

\begin{figure}[h]
\includegraphics[width=220 pt]{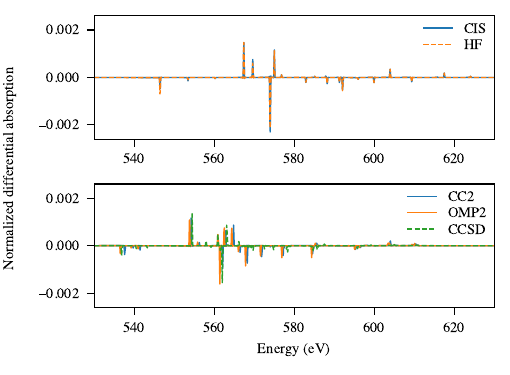}
    \caption{\label{fig:circdich_h2o2_methods_2}
    Differential spectra obtained with the cc-pVDZ basis set in the K-edge region of H$_2$O$_2$ with propagation direction along the $z$-axis.
    }
\end{figure}

The resulting CD spectra are plotted in Figs.~\ref{fig:circdich_h2o2_methods_0} and \ref{fig:circdich_h2o2_methods_2}.
As in the Ti$^{4+}$ simulations above, we see that the TDCIS and TDHF methods produce nearly identical CD spectra
with minor visual differences. The TDCC2 and TDOMP2 methods also yield similar CD spectra, producing the same sign pattern of the differential absorption peaks,
although the TDOMP2 peak positions are slightly more red-shifted than the TDCC2 ones relative to the TDHF peaks. The intensities of the TDCC2 and TDOMP2
spectra are significantly reduced compared with the TDHF and TDCIS spectra.
The TDCCSD method shifts the transition frequencies somewhat but produces an intensity of the dominant peak around $561$-$562\,\text{eV}$
which is closer to that of TDHF theory than the TDCC2 and TDOMP2 methods. Although this may indicate that high-level electron-correlation treatment is important, the deviation may also be caused by limited frequency resolution (see below).
Of course, the choice of carrier frequency will affect the relative peak magnitudes but further tests have shown that this effect is rather marginal
as long as $\omega$ is reasonably close to the transition energies.

\begin{figure}[h]
\includegraphics[width=220 pt]{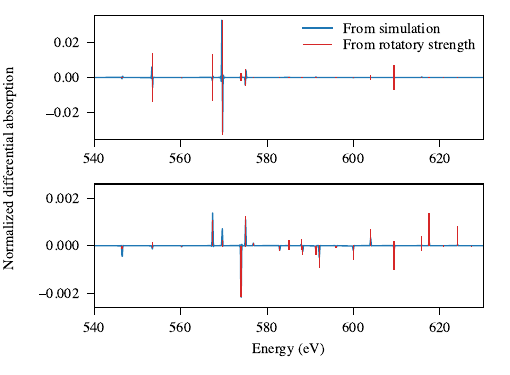}
    \caption{\label{fig:circdich_h2o2_with_cd_from_rot_str}
    Differential spectra obtained with the cc-pVDZ basis set in the K-edge region of H$_2$O$_2$ with propagation direction along the $x$ (top) and $z$ (bottom) axes, along with circular dichroism calculated from rotatory strengths.
    }
\end{figure}

\begin{figure}[h]
\includegraphics[width=220 pt]{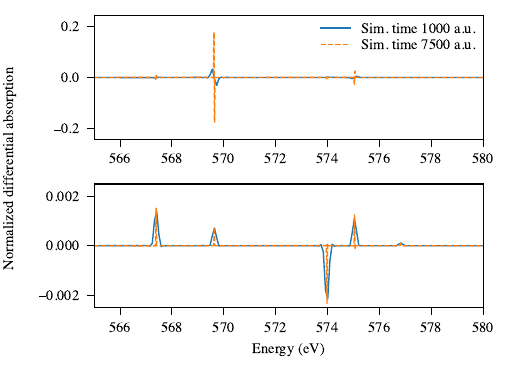}
    \caption{\label{fig:circdich_h2o2_convergence}
    TDCIS differential spectra obtained with the cc-pVDZ basis set in the K-edge region of H$_2$O$_2$ with propagation direction along the $x$ (top) and $z$ (bottom) axes, with simulation times of $1000\,\text{a.u.}$ and $7500\,\text{a.u.}$
    }
\end{figure}

Figure \ref{fig:circdich_h2o2_with_cd_from_rot_str} shows the CD spectrum obtained from the TDCIS simulations along with a stick spectrum calculated
from the rotatory strength tensors~\cite{PEDERSENTB1995ACAD} computed by full diagonalization of the CIS Hamiltonian matrix.
For both propagation directions, the stick spectrum is normalized such that the maximum peak is equal to the maximum peak from the corresponding TDCIS simulation.
Since the carrier frequency is $544.23\,\text{eV}$, it is expected that the peaks of the stick spectrum are smaller than the simulated peaks to the immediate left of the
dominant peak, and larger further to the right of the dominant peak. This is indeed what we observe
in the bottom panel of Fig. \ref{fig:circdich_h2o2_with_cd_from_rot_str}.
In the top panel, however, this is not the case. This can be ascribed to insufficient convergence.
The excited states of H$_2$O$_2$ in the C$_2$ geometry come in pairs, typically separated by 0.01 eV or less, formed by the lowering of symmetry relative to a 
planar, achiral (cis or trans) structure. 
For propagation in the $x$-direction, the CD for these pairs of states are of about the same magnitude but with opposite signs, causing lowering of the peak intensities.
Figure \ref{fig:circdich_h2o2_convergence} shows the effect of increasing the simulation time from $1000\,\text{a.u.}$ to $7500\,\text{a.u.}$ The change in the
bottom panel is relatively minor, while the dominant peak in the top panel has increased by an order of magnitude.
This is closer to the expected difference calculated from rotatory strength tensors. However, the peak at $567\,\text{eV}$ is still much suppressed,
which is caused by the states only being separated by about $0.0088\,\text{eV}$.

An overview of the occupied orbitals and the $11$ lowest-lying virtual orbitals is given in Table \ref{tab:h2o2_orbitals}.
\begin{table}[h]
\caption{Occupied orbitals and the $11$ lowest-lying virtual orbitals of H$_2$O$_2$ with the cc-pVDZ basis set. The orbital types are based on similarity with the peroxide ion.
    Due to the reduced symmetry, all orbitals are somewhat mixed.}
\label{tab:h2o2_orbitals}
\begin{tabular}{@{}llll@{}}
\toprule
No. & Energy (a.u.)       & \begin{tabular}[c]{@{}l@{}}Symmetry\\ label\end{tabular} & Type                      \\ \midrule
1   & -561.27036  & 1A                                                       & $\sigma$                  \\
2   & -561.26301  & 1B                                                       & $\sigma^*$                \\
3   & -39.992527  & 2A                                                       & $\sigma$                  \\
4   & -33.108905 & 2B                                                       & $\sigma$                  \\
5   & -19.465831   & 3B                                                       & $\pi_z$                   \\
6   & -18.833153 & 3A                                                       & $\pi_y$                   \\
7   & -16.202217 & 4A                                                       & $\sigma_{p_x}$            \\
8   & -14.251659  & 5A                                                       & $\pi_z^*$                 \\
9   & -13.038506 & 4B                                                       & $\pi_y^*$                 \\ \midrule
10  & 5.1526056   & 6A                                                       & $\sigma$                  \\
11  & 5.2866155   & 5B                                                       & $\sigma^*$                \\
12  & 7.7990137   & 6B                                                       & $\sigma_{p_x}^*$          \\
13  & 22.642558  & 7A                                                       & Mixed, dominant weight on H         \\
14  & 22.835716   & 7B                                                       & Mixed, dominant weight on H         \\
15  & 30.276955  & 8B                                                       & $\sigma^*_{s/p_x}$        \\
16  & 30.673602  & 8A                                                       & $\pi_y$/$\pi_z^*$         \\
17  & 31.260347  & 9A                                                       & $\pi_z^*$                 \\
18  & 32.689754    & 9B                                                       & $\sigma^*_{p_x}$/$pi_y^*$ \\
19  & 34.770868  & 10B                                                      & Mixed $\sigma$/$\pi$      \\
20  & 36.906177   & 10A                                                      & $\sigma^*_{s/p_x}$       
\end{tabular}
\end{table}
The core orbitals, 1$\sigma_s$ and 1$\sigma_s^*$, are separated by $7.3483\times 10^{-3}~\text{eV}$ and, hence, excitations from either of the core orbitals to
low-lying virtual orbitals will fall in the K pre-edge region.
The TDCIS spectrum contains $5$ main peaks below $580\,\text{eV}$ along with three smaller ones at $553.44\,\text{eV}$, $560.24\,\text{eV}$, and $576.82\,\text{eV}$.
The first peak at $546.48\,\text{eV}$ can be viewed as a transition to virtual orbitals 5B and 6B.
The main peak in Fig.~\ref{fig:circdich_h2o2_methods_0} at $569.63\,\text{eV}$ contains significant excitations to orbitals 8A, 9A and 9B,
which are orbitals with significant $\pi$ character, and with electron density mostly located on the oxygen atoms.
The main peak in Fig.~\ref{fig:circdich_h2o2_methods_2} at $573.97\,\text{eV}$ is mainly due to excitations to the 7B and 8B (and somewhat to 10A) orbitals.

\begin{figure}[h]
\includegraphics[width=220 pt]{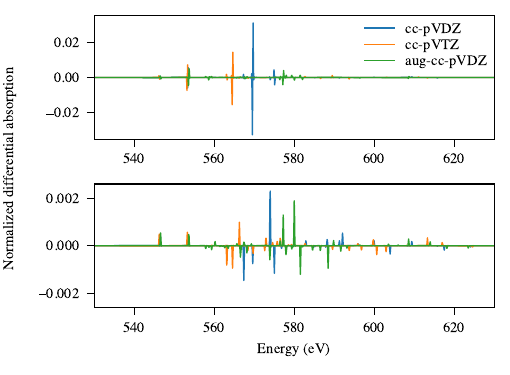}
    \caption{\label{fig:circdich_h2o2_basis}
    TDCIS CD spectra in the K-edge region of H$_2$O$_2$ with propagation direction along the $x$-axis (top) and $z$-axis (bottom).
    }
\end{figure}

Finally, noting that the cc-pVDZ basis set is insufficient for accurate predictions of CD spectra in general---see, e.g., Ref.~\citenum{Rizzo2006}---
we compare the TDCIS spectra with those obtained with larger basis sets in Fig.~\ref{fig:circdich_h2o2_basis}.
As expected, the basis-set effect is significant.
Going from double-zeta to triple-zeta basis retains some of the main features but the energies are red-shifted, whereas
the inclusion of diffuse orbitals in the aug-cc-pVDZ basis set leads to a much more radical change of the underlying dynamics due to a higher density of
excited states in the energy region around the carrier frequency.
More accurate predictions of transient CD spectra, especially with the higher-level TDCC methods, clearly require larger basis sets including diffuse functions.

\section{Concluding remarks}

We have derived a gauge invariant expression for the spectral response function which is applicable to transient absorption and emission spectra.
This expression is applicable both within and beyond the electric-dipole approximation. Using an enveloped plane-wave vector potential to formulate
the semiclassical matter-field interaction operator, simulations of laser-driven many-electron dynamics with a fixed atom-centered Gaussian basis set
can be straightforwardly carried out with no additional cost compared with the analogous electric-dipole simulations.
Numerical experiments show that beyond-dipole effects are fully captured without explicit multipole expansions, and that electric-dipole results are
correctly reproduced in the long wavelength limit. Circular (or, more general, elliptical) polarization is easily handled, as illustrated by 
preliminary simulations of anisotropic transient X-ray circular dichroism spectra.

Aimed at electronic ground and bound excited states, fixed atom-centered Gaussian basis sets do not support electronic continuum states and, consequently,
we have only considered low-intensity laser fields in this work. We are currently extending the approach presented here to more flexible bases that allow us
to study highly non-linear processes such as core ionization where the magnetic component of the electromagnetic field may play a decisive role.

\section{Supplementary material}
The supplementary material contains the molecular geometries of LiH, TiCl$_4$, and H$_2$O$_2$, and a brief analysis of the differences
between pulses with envelopes defined on the electric field and on the vector potential.

\section*{Acknowledgment}

This work was supported by the Research Council of Norway through its Centres of Excellence scheme, project number 262695.
The calculations were performed on resources provided by Sigma2---the National Infrastructure for High Performance Computing and
Data Storage in Norway, Grant No.~NN4654K.
SK and TBP acknowledge the support of the Centre for Advanced Study in Oslo, Norway, which funded and hosted the
CAS research project \emph{Attosecond Quantum Dynamics Beyond the Born-Oppenheimer Approximation} during the academic year
2021-2022. RL acknowledges the Swedish Research Council (VR, Grant No. 2020-03182) for funding.

\bibliography{main}

\end{document}


\preprint{APS/123-QED}

\title{Supplementary material for "Transient spectroscopy from time-dependent electronic-structure theory without multipole expansions"}

\author{Einar Aurbakken}
\email{einar.aurbakken@kjemi.uio.no}
\affiliation{Hylleraas Centre for Quantum Molecular Sciences,
Department of Chemistry, University of Oslo, Norway
}%

\author{Benedicte Sverdrup Ofstad}
\affiliation{Hylleraas Centre for Quantum Molecular Sciences,
Department of Chemistry, University of Oslo, Norway
}%

\author{H{\aa}kon Emil Kristiansen}
\affiliation{Hylleraas Centre for Quantum Molecular Sciences,
Department of Chemistry, University of Oslo, Norway
}%

\author{{\O}yvind Sigmundson Sch{\o}yen}
\affiliation{Department of Physics,
University of Oslo, Norway
}%

\author{Simen Kvaal}
\affiliation{Hylleraas Centre for Quantum Molecular Sciences,
Department of Chemistry, University of Oslo, Norway
}%

\author{Lasse Kragh S{\o}rensen}
\affiliation{University Library, University of Southern Denmark, DK-5230 Odense M, Denmark}%

\author{Roland Lindh}
\affiliation{Department of Chemistry---BMC,
Uppsala University, Sweden
}%

\author{Thomas Bondo Pedersen}
\email{t.b.pedersen@kjemi.uio.no}
\affiliation{Hylleraas Centre for Quantum Molecular Sciences,
Department of Chemistry, University of Oslo, Norway
}

\date{\today}

\maketitle

\section{Molecular geometries}
The nuclear Cartesian coordinates in Bohr for the molecules studied in the main manuscript are listed 
in Table \ref{tab:coord} below.
\begin{table}[h]
\caption{Cartesian coordinates in Bohr\label{tab:coord}}
\begin{tabular}{@{}lllll@{}}
\toprule
LiH & Li & 0.0 & 0.0 & 0.0 \\
    & H  & 0.0 & 0.0 & -3.0139491027559635 \\
\hline
TiCl$_4$ & Ti & 0.0 & 0.0 & 0.0 \\
         & Cl & 0.0 & 0.0 & 4.1007056904379215 \\
         & Cl & 3.86618240181189 & 0.0 & -1.3669018968126405 \\
         & Cl &-1.93309120090594 & 3.348212175633433 &-1.3669018968126405 \\
         & Cl &-1.93309120090594 &-3.348212175633433 &-1.3669018968126405 \\
\hline
H$_2$O$_2$ & O & 1.3936730169115057 & 0.0 & 0.0 \\
           & O &-1.3936730169115057 & 0.0 & 0.0 \\
           & H & 1.5439062438192541 & 0.8972419639723157 & 1.5476856960685057 \\
           & H &-1.5439062438192541 & 0.8972419639723157 &-1.5476856960685057 \\
\toprule
\end{tabular}
\end{table}

\section{Pulse shapes and frequency components}
\subsection{Definitions of pulses}

\begin{figure}[h]
\includegraphics[width=\textwidth]{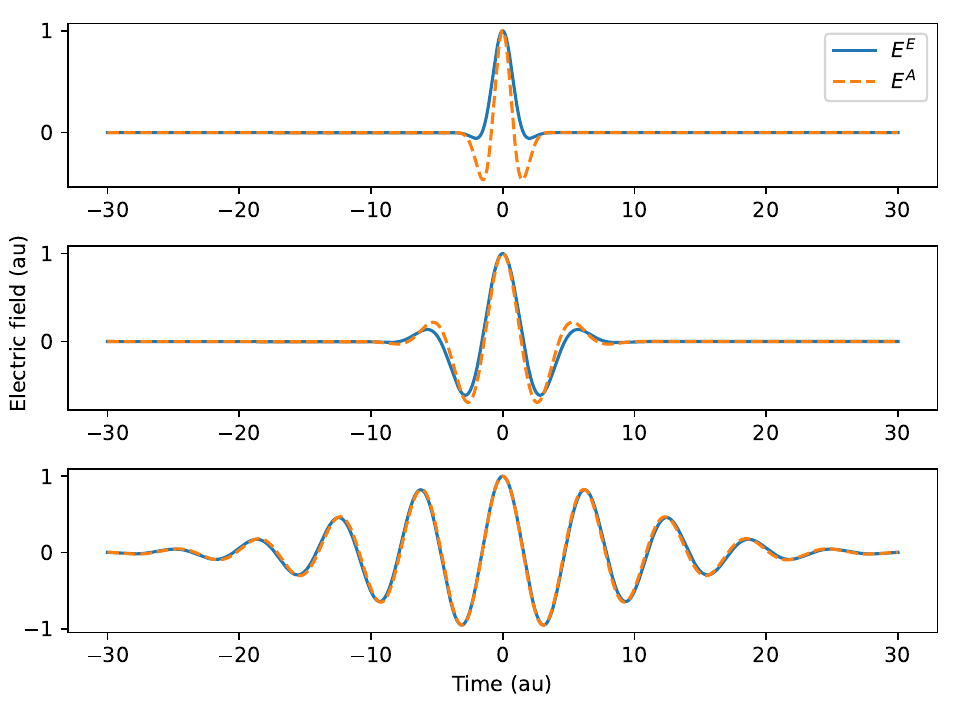}
\caption{\label{fig:three_sigma_values}Comparison of $E^E$ (envelope on the electric field) and $E^A$ (envelope on the vector potential). Common parameters were $\omega=1$, $\phi=0$. Top: $\sigma=1$, middle: $\sigma=3$, bottom: $\sigma=10$.}
\end{figure}

Electric field pulse shapes of the form 
\begin{align}\label{eq:EsupE}
    \bm{E}^E(t) = \sum_{n=1}^N \bm{E}^E_n(t) = \sum_{n=1}^N E_n \bm{u}_n \cos{\left[\omega_n(t-t_n) - \phi_n \right]} G_n(t)
\end{align}
are common in the literature (usually on singular form, $\bm{E}^E(t)=\bm{E}^E_1(t)$). In Eq. (\ref{eq:EsupE}), $n$ is the pulse number, $N$ is the number of pulses, $E_n$ is the maximum field strength, $\bm{u}_n$ is a real polarization vector, $\omega_n$ is the carrier frequency, $t_n$ the central time, $G_n(t)$ is the envelope and $\phi_n$ is the carrier-envelope phase. Superscript $E$ signifies that the envelope has been placed on the electric field. Here we will use Gaussian envelopes defined by
\begin{align}
    G_n(t) = \exp\left[\frac{-(t-t_n)^2}{2\sigma_n^2}\right]
\end{align}
where $\sigma_n$ is the standard deviation. Alternatively, we can define the electric field by the vector potential in velocity gauge, $\bm{E}(t)=-\partial_t \bm{A}(t)$, where we define the vector potential as
\begin{align}
    \bm{A}(t) = -\sum_{n=1}^N A_n \bm{u}_n \sin{\left[\omega_n(t-t_n) - \phi_n \right]} G_n(t)
\end{align}
The corresponding electric field is then
\begin{align}\label{eq:EsupA}
    \bm{E}^A(t) &= \sum_{n=1}^N \bm{E}^A_n(t) \nonumber \\ &= \sum_{n=1}^N E_n \bm{u}_n \left\{\cos{\left[\omega_n(t-t_n) - \phi_n \right]} -\frac{t-t_n}{\omega_n \sigma_n^2} \sin{\left[\omega_n(t-t_n) - \phi_n \right]} \right\} G_n(t)
\end{align}
where superscript $A$ denotes that the envelope was placed on the vector potential and $E_n=\omega_n A_n$. For a given carrier-frequency $\omega_n$, $E^A_n(t)$ will converge to $E^E_n(t)$ for increasing values of $\sigma_n$, but in cases where $\sigma_n$ is small in the sense that we have effectively few cycles, $E^A_n(t)$ may differ significantly from $E^E_n(t)$. An illustration of this is shown in Figure \ref{fig:three_sigma_values}.

\subsection{Frequency components}

The zero-frequency (dc) component of the electric field is given by
\begin{align}
    \Tilde{\bm{E}}(0) = \int_{-\infty}^{\infty} \bm{E}(t) dt
\end{align}
 The dc component should vanish in the far field approximation of the Maxwell equations \cite{Rauch2006}. This condition is by definition satisfied by the pulse given in Eq. (\ref{eq:EsupA}). It is in general not satisfied for pulses defined by Eq. (\ref{eq:EsupE}), but the dc component will vanish for phases $\phi_n = (m+1/2)\pi$, $m\in \mathbb{Z}$,  since the function is then odd.

In what follows, we use the magnitude of the Fourier component defined as
\begin{align}\label{eq:fourier_magnitude}
    |\Tilde{f}(\omega)| = \left|\int_{\infty}^{\infty}f(t)e^{-i\omega t}dt \right|
\end{align}
to analyse the frequency components in the pulses.

\subsection{Example: LiH pump-probe}

\begin{figure}[h]
\includegraphics[width=\textwidth]{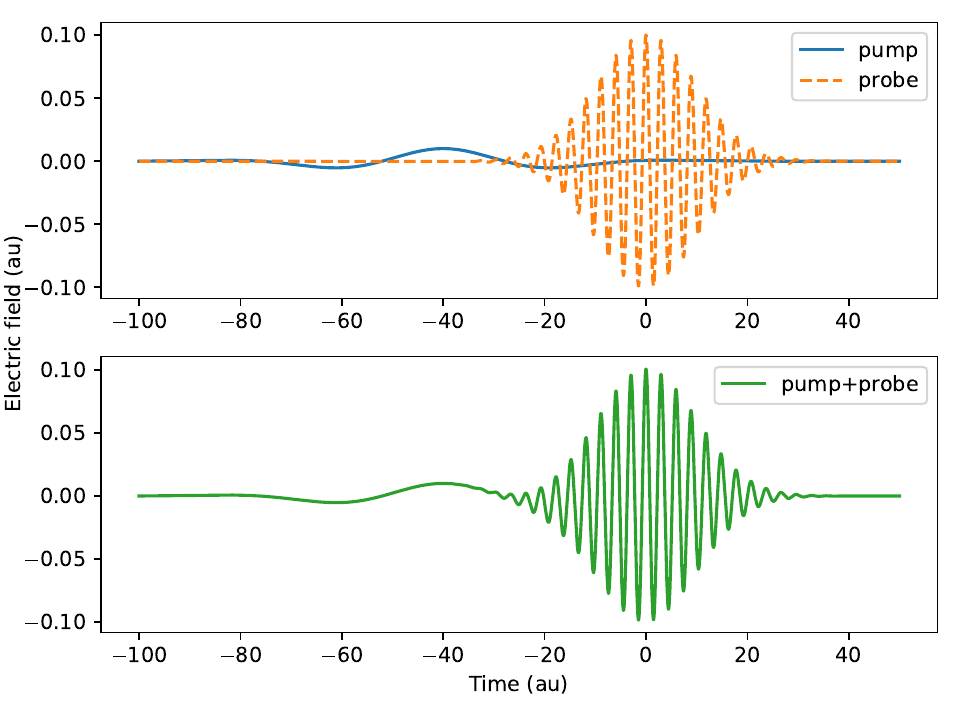}
\caption{\label{fig:len_E_pump_probe}Illustration of the electric field of the LiH pump-probe simulations using $E^E$.}
\end{figure}

We will here look at the pump-probe pulse used for LiH in Ref. \cite{SkeidsvollAndreasS2020Tctf}, where the electric field was on the form given in Eq. {\ref{eq:EsupE}} and polarized in the z-direction. An illustration of the pulses is given in Figure \ref{fig:len_E_pump_probe}. The parameters are given in Table \ref{tab:params}.

\begin{table}[H]
\begin{center}
\caption{Parameters used in pump-probe setup. All values are in atomic units.}
\label{tab:params}
\begin{tabular}{@{}lll@{}}
\toprule
Parameters & Pump     & Probe    \\ \midrule
$E_n$      & 0.01     & 0.1      \\
$\omega_n$ & 0.130551 & 2.118698 \\
$\phi_n$   & 0        & 0        \\
$\sigma_n$ & 20       & 10       \\
$t_n$      & -40      & 0        \\ \bottomrule
\end{tabular}
\end{center}
\end{table}

\begin{figure}[h]
\includegraphics[width=\textwidth]{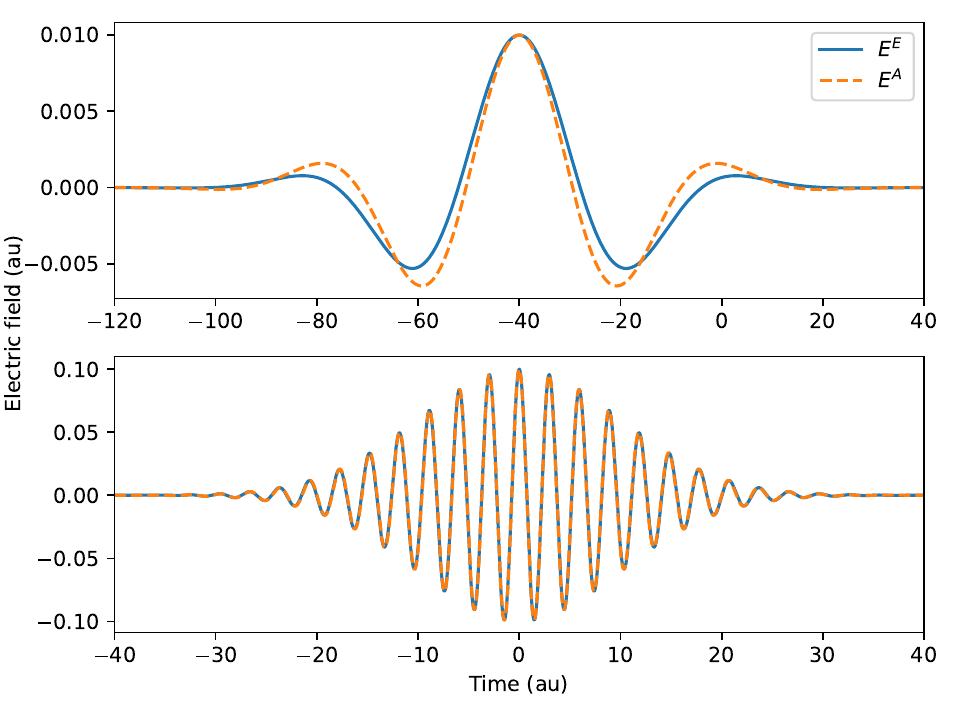}
\caption{\label{fig:len_E_vs_A_pump_probe}Comparison of the pump and probe pulses using $E^E$ and $E^A$ pulses.}
\end{figure}

Here, we will refer to these generally as $E^E$ pulses, as opposed to the corresponding pulses generated with Eq. (\ref{eq:EsupA}) using the same parameters, which we will refer to as $E^A$ pulses. A comparison between the $E^E$ and $E^A$ pump and probe pulses is shown in Figure \ref{fig:len_E_vs_A_pump_probe}. As expected, the difference between the probe pulses is quite small since they have (relatively) many cycles, while the difference between the pump pulses is quite significant since they have few cycles.

\begin{figure}[h]
\includegraphics[width=\textwidth]{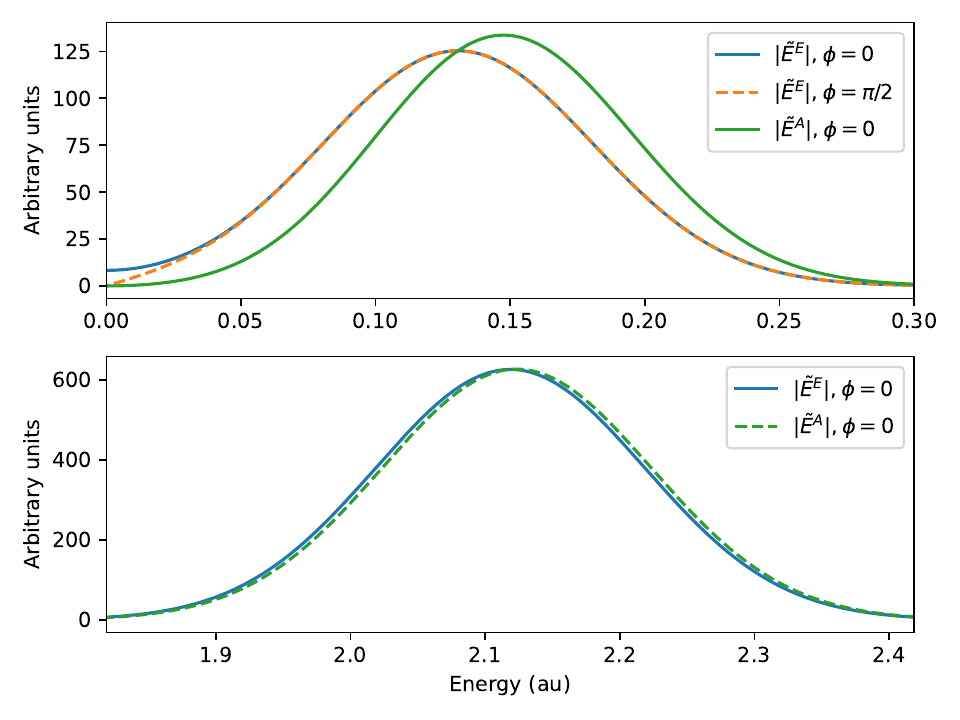}
\caption{\label{fig:fourier_magnitudes}Magnitude of the Fourier components in the pump (top) and probe (bottom) pulses using the $E^A$ pulse and $E^E$ pulses with carrier-envelope phase $0$ and $\pi/2$.}
\end{figure}

Figure \ref{fig:fourier_magnitudes} shows the magnitude of Fourier components of the $E^E$ and $E^A$ pulses. In the figure, an $E^E$ pump pulse with $\phi=\pi/2$ has also been included. We note that $|\Tilde{E}^E(\omega_n)|$ and $|\Tilde{E}^A(\omega_n)|$ are similar, but the central frequencies of the $E^A$ pulses are higher, and have larger maximum Fourier components. Expectedly, these differences are quite large for the pump pulses and much smaller for the probe pulses. Also, we see (as mentioned above) that the $E^E$ pulse with $\phi=0$ has a non-vanishing zero-frequency component.

\begin{figure}[h]
\includegraphics[width=\textwidth]{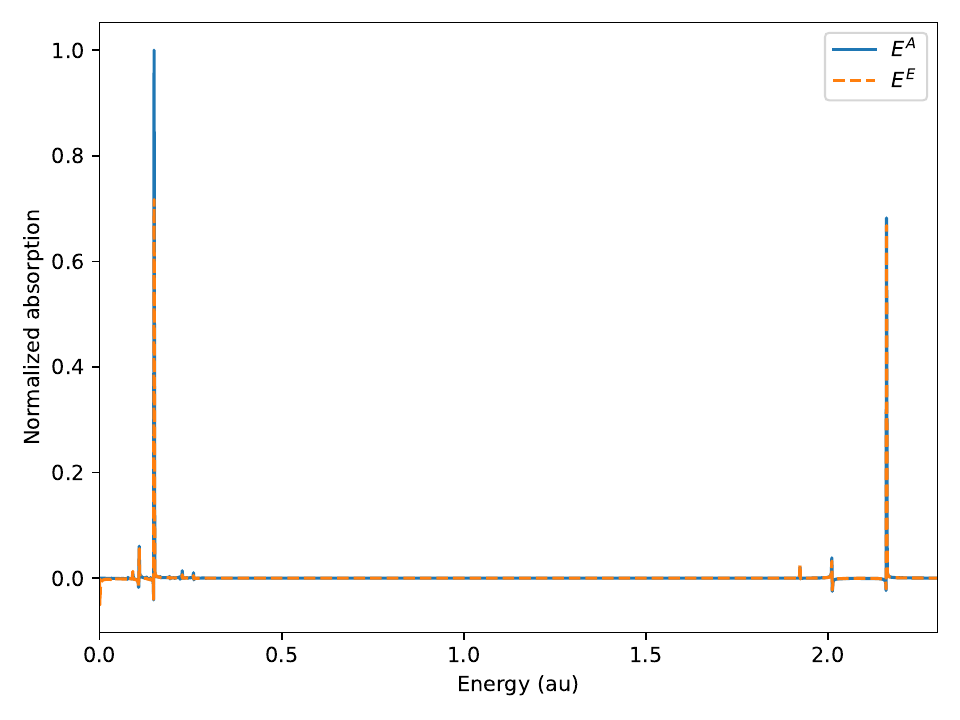}
\caption{\label{fig:spectrum_len_E_vs_A}Transient spectrum using the $E^A$ and $E^E$ pulses.}
\end{figure}

Figure \ref{fig:spectrum_len_E_vs_A} shows a comparison between a pump-probe spectrum generated with the $E^E$ and corresponding $E^A$ pulses using TDRCIS. We see that the spectra show very small deviations in the "probe region", while the difference in absorption in the "pump region" is large, in accordance with the preceding analysis.

\begin{figure}[h]
\includegraphics[width=\textwidth]{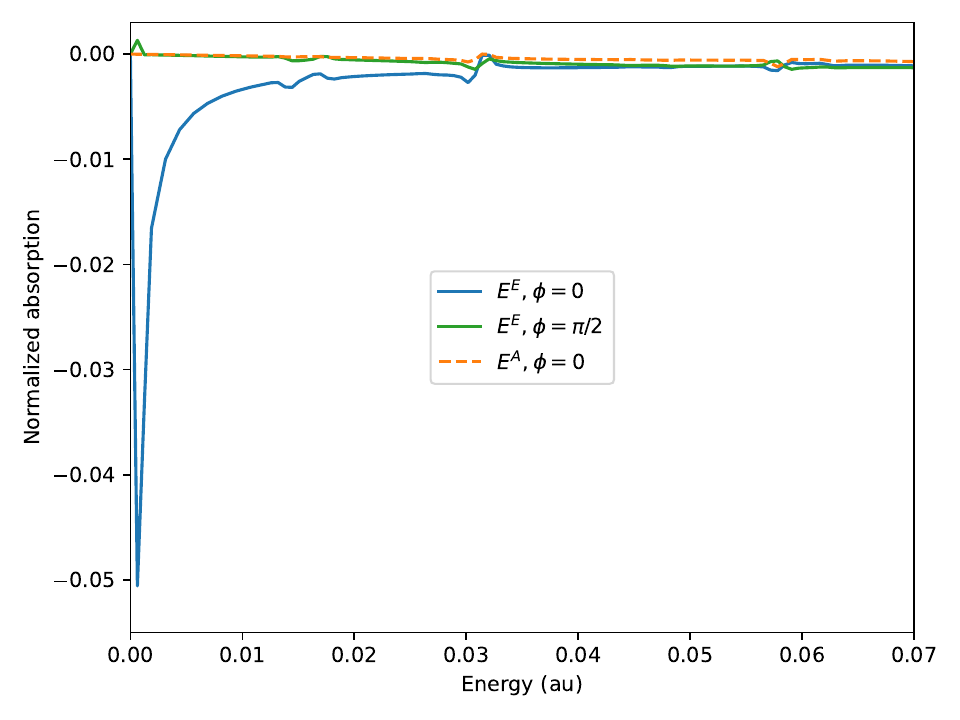}
\caption{\label{fig:spectrum_zero_frequency}Zero frequency region of transient spectrum using the $E^A$ pulse and $E^E$ pulses with carrier-envelope phase $0$ and $\pi/2$.}
\end{figure}

We can also note that the $E^E$ spectrum has an unphysical close-to zero frequency component. This is also visible in Fig. 1 in Ref. \cite{SkeidsvollAndreasS2020Tctf}. Figure \ref{fig:spectrum_zero_frequency} shows this in in more detail, and also includes an $E^E$ spectrum with $\phi_1 = \phi_2 = \pi/2$. The zero-frequency component is practically gone in the spectra generated from electric fields with zero dc component, and it is therefore natural to connect this to the non-vanishing dc component of the electric field.

\bibliography{references}